\documentclass[fleqn,10pt]{wlscirep}
\usepackage{amsmath,amsfonts,amssymb,graphics,graphicx,epsfig,color,times,bbm,subfigure,hyperref,keyval}

\newcommand{\ket}[1]{|#1\rangle}
\newcommand{\bra}[1]{\langle #1|}
\newcommand{\proj}[1]{\ket{#1}\bra{#1}}
\newcommand{\beq}{\begin{equation}}
\newcommand{\eeq}{\end{equation}}
\newcommand{\beqa}{\begin{eqnarray}}
\newcommand{\eeqa}{\end{eqnarray}}

\newcommand{\nc}{\newcommand}

\nc{\cA}{{\cal A}} \nc{\cB}{{\cal B}} \nc{\cC}{{\cal C}}
\nc{\cD}{{\cal D}} \nc{\cE}{{\cal E}} \nc{\cF}{{\cal F}}
\nc{\cG}{{\cal G}} \nc{\cH}{{\cal H}} \nc{\cI}{{\cal I}}
\nc{\cJ}{{\cal J}} \nc{\cK}{{\cal K}} \nc{\cL}{{\cal L}}
\nc{\cM}{{\cal M}} \nc{\cN}{{\cal N}} \nc{\cO}{{\cal O}}
\nc{\cP}{{\cal P}} \nc{\cQ}{{\cal Q}} \nc{\cR}{{\cal R}}
\nc{\cS}{{\cal S}} \nc{\cT}{{\cal T}} \nc{\cU}{{\cal U}}
\nc{\cV}{{\cal V}} \nc{\cW}{{\cal W}} \nc{\cX}{{\cal X}}
\nc{\cZ}{{\cal Z}}

\newcommand{\abs}[1]{|#1|}

\newcommand{\braket}[2]{\langle#1|#2\rangle}

\newcommand{\Gl}{{\text{G}}_{\text{l}}}

\newcommand{\Gcl}{{\text{G}}^{\text{c}}_{\text{l}}}

\def\squareforqed{\hbox{\rlap{$\sqcap$}$\sqcup$}}
\def\qed{\ifmmode\squareforqed\else{\unskip\nobreak\hfil
\penalty50\hskip1em\null\nobreak\hfil\squareforqed
\parfillskip=0pt\finalhyphendemerits=0\endgraf}\fi}
\def\endenv{\ifmmode\;\else{\unskip\nobreak\hfil
\penalty50\hskip1em\null\nobreak\hfil\;
\parfillskip=0pt\finalhyphendemerits=0\endgraf}\fi}
\newenvironment{proof}{\noindent \textbf{{Proof.~} }}{\qed}
\def\Dbar{\leavevmode\lower.6ex\hbox to 0pt
{\hskip-.23ex\accent"16\hss}D}
\makeatletter
\def\url@leostyle{%
  \@ifundefined{selectfont}{\def\UrlFont{\sf}}{\def\UrlFont{\small\ttfamily}}}
\makeatother
\usepackage{hyperref,color}
\def\bea{\begin{eqnarray}}
\def\eea{\end{eqnarray}}
\def\bpf{\begin{proof}}
\def\epf{\end{proof}}
\def\tr{{\rm Tr}}

\def\a{\alpha}
\def\b{\beta}

\def\m{\mu}

\def\r{\rho}

\def\ph{\varphi}

\def\ps{\psi}

\title{Lower and upper bounds for entanglement of R\'{e}nyi-$\alpha$ entropy}

\author[1,$^\ast$]{Wei Song}

\author[2,3,$^\dag$]{Lin Chen}

\author[1]{Zhuo-Liang Cao}
\affil[1]{Institute for Quantum Control and Quantum Information, and School
of Electronic and Information Engineering, Hefei Normal University, Hefei 230601, China}
\affil[2]{School of Mathematics and Systems Science, Beihang University, Beijing 100191, China}
\affil[3]{International Research Institute for Multidisciplinary Science, Beihang University, Beijing 100191, China}

\affil[$^\ast$]{wsong1@mail.ustc.edu.cn}
\affil[$^\dag$]{linchen@buaa.edu.cn}

\begin{abstract}
Entanglement R\'{e}nyi-$\alpha$ entropy is an entanglement measure. It reduces to the standard entanglement of formation when $\alpha$ tends to 1. We derive analytical lower and upper bounds for the entanglement R\'{e}nyi-$\alpha$ entropy of arbitrary dimensional bipartite quantum systems. We also demonstrate the application our bound for some concrete examples. Moreover, we establish the relation between entanglement R\'{e}nyi-$\alpha$ entropy and some other entanglement measures.
\end{abstract}

\begin{document}
\flushbottom
\maketitle

\section*{Introduction}

Quantum entanglement is one the most remarkable features of quantum mechanics and is the
key resource central to much of quantum information applications. For this reason, the characterization and quantification of entanglement has become an important problem in quantum-information science \cite{Neilsen00}. A number of entanglement measures have been proposed for bipartite states such as the entanglement of formation (EOF) \cite{Bennett96pra}, concurrence \cite{woo98prl}, relative entropy \cite{Vedral97prl}, geometric entanglement \cite{Wei03pra}, negativity \cite{Vidal02pra} and squashed entanglement  \cite{Christandl03jmp,Yang09ieee}. Among them EOF is one of the most famous measures of entanglement. For a pure bipartite state $\left| \psi \right\rangle_{AB}$ in the Hilbert space, the EOF is given by
\beq\label{q1}
E_F\left( {\left| \psi  \right\rangle_{AB}} \right) = S\left( {\rho _A } \right),
\eeq
where $S\left( \rho_A  \right) :=  - \tr\rho_A \log \rho_A$ is the von Neumann entropy of the reduced density operator of system $A$. Here ``log'' refers to the logarithm of base two. The situation for bipartite mixed states $\rho_{AB}$ is defined by the convex roof
\beq\label{q2}
E_F \left( \rho_{AB}  \right) = \min \sum\limits_i {p_i E_F \left( {\left| {\psi _i } \right\rangle_{AB} } \right)},
\eeq
where the minimum is taken over all possible pure state decompositions of ${\rho_{AB}=\sum\limits_i{p_i
\left| {\psi _i } \right\rangle _{AB} \left\langle {\psi _i } \right|} }$ with $\sum\limits_i {p_i }  = 1$ and $p_i>0$. The EOF provides an upper bound on the rate at which maximally
entangled states can be distilled from $\rho$ and a lower
bound on the rate at which maximally entangled states needed
to prepare copies of $\rho$ \cite{Hayden01jpa}. For two-qubit systems, an elegant formula for EOF
was derived by Wootters in \cite{woo98prl}. However, for the general
highly dimensional case, the evaluation of EOF remains a nontrivial
task due to the the difficulties in minimization procedures\cite{Huang14njp}. At
present, there are only a few analytic formulas for EOF including the
isotropic states \cite{Terhal00prl}, Werner states \cite{Vollbrecht01pra} and Gaussian
states with certain symmetries \cite{Giedke03prl}. In order to evaluate the entanglement measures, many efforts have also been devoted to the study of lower
and upper bounds of different entanglement measures \cite{Vidal02prl,Fei04rmp,Gerjuoy03pra,Mintert04prl,Chen05prl01,Chen05prl02,Osborne05pra,Mintert05prl,Fei06pra,Datta07pra,Mintert07prl,Zhang08pra,Ma10pra,Li11pra,Zhao11pra,Sabour12pra,Chen12prl,Nicacio14prl,Nicacio14pra}. Especially, Chen \emph{et al} \cite{Chen05prl01} derived an analytic
lower bound of EOF for an arbitrary bipartite mixed state, which established a
bridge between EOF and two strong separability criteria. Based on this idea, there are several improved lower and upper bounds for EOF presented in \cite{Li10pra,Zhang11pra,Zhu12pra,Zhang15}. While the entanglement of formation is the most common measure of entanglement, it is not the unique measure. There are other measures such as entanglement R\'{e}nyi-$\alpha$ entropy (ER$\alpha$E) which is the generalization of the entanglement of formation. The ER$\alpha$E has a continuous spectrum parametrized by the non-negative real parameter $\alpha$. For a bipartite pure state $\left| \psi  \right\rangle _{AB}$,
the ER$\alpha$E is defined as
\cite{Kim10jpa}
\beq\label{q3}
E_{\alpha}(\left| \psi  \right\rangle _{AB}):= S_\alpha(\rho_A) := \frac{1}{1-\alpha}\log (\mbox{tr}\rho _A^\alpha),
\eeq
where $S_\alpha(\rho_A)$ is the R\'{e}nyi-$\alpha$ entropy.
Let $\mu_1,\cdots,\mu_m$ be the eigenvalues of the reduced density matrix $\rho_A$ of $\left| \psi  \right\rangle _{AB}$. We have
\beq
S_\alpha(\rho_A)={1\over 1-\alpha}\log(\sum_i\mu_i^{\alpha})
:=H_\alpha  \left( {\vec \mu } \right),
\eeq
where ${\vec \mu }$ is called the Schmidt vector $\left( {\mu _1 ,\mu _2 , \cdots ,\mu _m } \right)$.
The R\'{e}nyi-$\alpha$ entropy is additive on independent states and has found important applications in characterizing quantum phases with
differing  computational power \cite{Cui12nc}, ground state properties
in many-body systems \cite{Franchini14prx}, and topologically ordered states
\cite{Flammia09prl,Halasz13prl}. Similar to the convex roof in \eqref{q2}, the ER$\alpha$E of a bipartite mixed state $\rho _{AB}$
is defined as
\beqa\label{q4}
E_\alpha(\rho _{AB})=\mbox{min} \sum_i p_i E_\alpha(\ket{\psi _i }_{AB}).
\eeqa
It is known that the R\'{e}nyi-$\alpha$ entropy converges to the von Neumann
entropy when $\alpha$ tends to 1. So the ER$\a$E reduces to the EOF when $\alpha$ tends to 1.
Further ER$\alpha$E is not increased under local operations and classical communications (LOCC) \cite{Kim10jpa}. So the ER$\a$E is an entanglement monontone, and becomes zero if and only if $\r_{AB}$ is a separable state.

An explicit expression of ER$\alpha$E has been derived for two-qubit mixed state with $\alpha\geq(\sqrt{7}-1)/2\simeq 0.823$ \cite{Kim10jpa,yxwang15arx}. Recently, Wang \emph{et al }\cite{yxwang15arx} further derived the analytical formula of ER$\alpha$E for Werner states and isotropic states. However, the general analytical results of ER$\a$E even for the two-qubit mixed state with arbitrary parameter $\alpha$ is still a challenging problem.

The aim of this paper is to provide computable lower and upper bounds for ER$\alpha$E of arbitrary dimensional bipartite quantum systems, and these results might be utilized to investigate the monogamy relation\cite{ckw00pra,Osborne06prl,Bai14prl,Bai14pra} in high-dimensional states. The key step of our work is to relate the lower or upper bounds with the concurrence which is relatively easier to dealt with. We also demonstrate the application of these bounds for some examples. Furthermore, we derive the relation of ER$\alpha$E with some other entanglement measures.

\section*{Lower and upper bounds for entanglement of R\'{e}nyi-$\alpha$ entropy}
\label{sec:bound}

For a bipartite pure state with Schmidt decomposition $\left| \psi  \right\rangle  = \sum\nolimits_{i = 1}^m {\sqrt {\mu _i } \left| {ii} \right\rangle }$, the concurrence of $\left| \psi  \right\rangle$ is given by $c(\ket{\ps}) := \sqrt{2(1-\tr \r_A^2)} = \sqrt {2\left( {1 - \sum\nolimits_{i = 1}^m {\mu _i^2 } } \right)}$. The expression $1-\tr\r_A^2$ is also known as the mixedness and linear entropy \cite{gks14,Chen14pra}.
The concurrence
of a bipartite mixed state $\rho$ is defined by the convex roof $c\left( \rho  \right) = \min \sum\limits_i {p_i c\left( {\left| {\psi _i } \right\rangle } \right)}$ for all possible pure state decompositions of ${\rho=\sum\limits_i{p_i\left| {\psi _i } \right\rangle\left\langle {\psi _i } \right|} }$. A series of lower and upper bounds for concurrence have been obtained in Refs  \cite{Chen05prl02,Mintert07prl,Zhang08pra}. For example, Chen \emph{et al} \cite{Chen05prl02} provides a lower bound for the concurrence by making a connection with the known strong separability criteria \cite{Peres96prl,Chen03qic}, i.e.,
\beqa\label{q5}
c\left( \rho  \right) \ge \sqrt {\frac{2}{{m\left( {m - 2} \right)}}} \left( {\max \left( {\left\| {\rho ^{T_A } } \right\|,\left\| {\mathcal{R}\left( \rho  \right)} \right\|} \right) - 1} \right),
\eeqa
for any $m \otimes n(m \le n)$ mixed quantum system. The $\|\cdot\|$ denotes the trace norm and $T_A$ denotes the partial transpose. Another important bound of squared concurrence used in our work is given by \cite{Mintert07prl,Zhang08pra}

\beqa\label{q6}
\tr\left( {\rho  \otimes \rho V_i } \right) \le \left[ {C\left( \rho  \right)} \right]^2  \le \tr\left( {\rho  \otimes \rho K_i } \right),
\eeqa
with $V_1  = 4( {P_ - ^{\left( 1 \right)}  - P_ + ^{\left( 1 \right)} } ) \otimes P_ - ^{\left( 2 \right)}$, $V_2  = 4P_ - ^{\left( 1 \right)}  \otimes ( {P_ - ^{\left( 2 \right)}  - P_ + ^{\left( 2 \right)} } )$, $K_1  = 4(P_ - ^{\left( 1 \right)}  \otimes I^{\left( 2 \right)})
$, $K_2  = 4( {I^{\left( 1 \right)}  \otimes P_ - ^{\left( 2 \right)} } )$ and $P_ - ^{\left( i \right)} (P_ + ^{\left( i \right)} )$ is the projector on the antisymmetric (symmetric) subspace of the two copies of the $i$th system. These bounds can be directly measured and can also be written as

\beqa\label{q7}
\tr\left( {\rho  \otimes \rho V_1 } \right) &=& 2\left( {\tr\rho ^2  - \tr\rho _A^2 } \right), \\
 \tr\left( {\rho  \otimes \rho V_2 } \right) &=& 2\left( {\tr\rho ^2  - \tr\rho _B^2 } \right), \\
 \tr\left( {\rho  \otimes \rho K_1 } \right) &=& 2\left( {1 - \tr\rho _A^2 } \right), \\
 \tr\left( {\rho  \otimes \rho K_2 } \right) &=& 2\left( {1 - \tr\rho _B^2 } \right).
\eeqa

Below we shall derive the lower and upper bounds of ER$\alpha$E based on these existing bounds of concurrence. Different states may have the same concurrence. Thus the value of $H_\alpha  \left( {\vec \mu } \right)$ varies with different Schmidt coefficients ${\mu _i}$ for fixed concurrence. We define two functions

\beqa\label{q8}
& &R_U ( c ) = \max \left\{ {H_\alpha  ( {\vec \mu } )|\sqrt {2( {1 - \sum\nolimits_{i = 1}^m {\mu _i^2 } } )}  \equiv c} \right\}, \\
& &R_L ( c ) = \min \left\{ {H_\alpha  ( {\vec \mu } )|\sqrt {2( {1 - \sum\nolimits_{i = 1}^m {\mu _i^2 } } )}  \equiv c} \right\}.
\eeqa
The derivation of them is equivalent to finding the maximal and minimal of $H_\alpha  \left( {\vec \mu } \right)$. Notice that the definition of $H_\alpha  \left( {\vec \mu } \right)$, it is equivalent to find the maximal and minimal of $\sum^m_{i=1} {\mu _i^\alpha  }$ under the constraint $\sqrt {2( {1 - \sum\nolimits_{i = 1}^m {\mu _i^2 } } )}  \equiv c$ since the logarithmic function is a monotonic function . With the method of Lagrange multipliers we obtain the necessary condition for the maximum and minimum of $\sum^m_{i=1} {\mu _i^\alpha  }$ as follows

\beqa\label{q14}
\alpha \mu _i^{\alpha  - 1}  = 2\lambda _1 \mu _i  - \lambda _2,
\eeqa
where $\lambda _1, \lambda _2$ denote the Lagrange multipliers. This equation has maximally two nonzero solutions $\gamma$ and $\delta$ for each $\mu _i$. Let $n_1$ be the number of entries where  $\mu _i=\gamma$ and $n_2$ be the number of entries where $\mu _i=\delta$. Thus the derivation is reduced to maximizes or minimizes the function
\beqa\label{q15}
R_{n_1 n_2 } \left( c \right)=\frac{1}{{1 - \alpha }}\log \left( {n_1 \gamma ^\alpha   + n_2 \delta ^\alpha  } \right),
\eeqa
under the constrains

\beqa\label{q16}
n_1 \gamma  + n_2 \delta  = 1,2\left( {1 - n_1 \gamma ^2  - n_2 \delta ^2 } \right) = c^2,
\eeqa
where $n_1  + n_2  = d \le m$. From Eq. \eqref{q16} we obtain two solutions of $\gamma$

\beqa\label{q17}
\gamma _{n_1 n_2 }^ \pm  = \frac{{n_1  \pm \sqrt {n_1^2  - n_1 ( {n_1  + n_2 } )[ {1 - n_2 ( {1 - c^2 /2} )} ]} }}{{n_1 ( {n_1  + n_2 } )}},
\eeqa

\beqa\label{q18}
\delta _{n_1 n_2 }^ \pm   = \frac{{1 - n_1 \gamma _{n_1 n_2 }^ \pm  }}{{n_2 }},
\eeqa
with $\max \{ \sqrt {2\left( {n_1  - 1} \right)/n_1 } ,\sqrt {2\left( {n_2  - 1} \right)/n_2 } \}  \le c \le \sqrt {2\left( {d - 1} \right)/d}$. Because $\gamma _{n_2 n_1 }^ -   = \delta _{n_1 n_2 }^ +  ,\delta _{n_2 n_1 }^ -   = \gamma _{n_1 n_2 }^ +$, we should only consider the case for $\gamma _{n_1 n_2 }^ +$. When $n_2=0$, $\gamma$ can be uniquely determined by the constrains thus we omit this case.

\begin{figure}[htb]
\includegraphics[scale=0.7,angle=0]{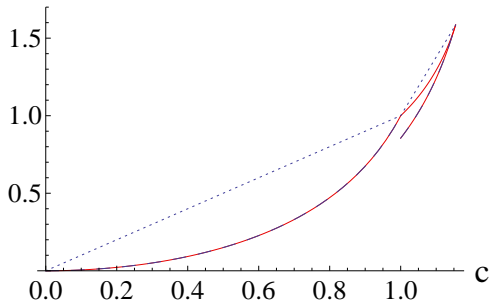}
\caption{(color online). The plot of lower bound (dashed line) and upper bound (dotted line) for $\alpha=3, m=3$. The upper bound consists of two segments and the lower bound consists of three segments. The solid line corresponds to $R_{11},R_{12}$ and $R_{21}$.}\label{fig1}
\end{figure}

When $m=3$, the solution of Eq.(\ref{q15}) is $R_{12}(c)$ and $R_{21}(c)$ for $1 < c \le 2/\sqrt 3$. After a direct calculation we find $R_{12}(c)$ and $R_{21}(c)$ are both monotonically
function of the concurrence $c$, and $R_{12}(2/\sqrt 3)=R_{21}(2/\sqrt 3)$. In order to compare the value of $R_{12}(c)$ and $R_{21}(c)$ we only need to compare the value of them at the endpoint $c=1$. For convenience we divide the problem into three cases. If $0 < \alpha <2$, then $R_{12}(1)>R_{21}(1)$; If $\alpha =2$, then $R_{12}(1)=R_{21}(1)$; If $\alpha >2$, then $R_{12}(1)<R_{21}(1)$. Thus we conclude that the maximal and minimal function of $H_\alpha  \left( {\vec \mu } \right)$ is given by $R_{21}(c)$ and $R_{12}(c)$ respectively for $\alpha  > 2$. When $\alpha  < 2$, the maximal and minimal function of $H_\alpha  \left( {\vec \mu } \right)$ is $R_{12}(c)$ and $R_{21}(c)$ respectively. When $\alpha  = 2$, we can check that the two functions $R_{21}(c)$ and $R_{12}(c)$ always have the same value for $1 < c \le 2/\sqrt 3$. In the general case for $m=d$, numerical
calculation shows the following results

(i) When $\alpha  > 2$,
\beqa\label{q10}
R_L ( c ) = \frac{{\log  [ {( {\gamma _{1,d - 1}^ +  } )^\alpha   + ( {d - 1} )^{1 - \alpha } ( {1 - \gamma _{1,d - 1}^ +  } )^\alpha  } ]}}{{1 - \alpha }},
\eeqa

\beqa\label{q11}
R_U ( c ) = \frac{{\log [ {( {\gamma _{1,d - 1}^ -  } )^\alpha   + ( {d - 1} )^{1 - \alpha } ( {1 - \gamma _{1,d - 1}^ -  } )^\alpha  } ]}}{{1 - \alpha }},
\eeqa
with $\sqrt {2( {d - 2} )/(d - 1)}  < c \le \sqrt {2(d - 1)/d}$, $1 \le d \le m - 1$ and $
\gamma _{1,d - 1}^ \pm   = ( {2 \pm \sqrt {2( {d - 1} )[ {d(2 - c^2 ) - 2} ]} } )/2d$.

(ii) When $\alpha  < 2$,
\beqa\label{q12}
R_L ( c ) = \frac{{\log [ {( {\gamma _{1,d - 1}^ -  } )^\alpha   + ( {d - 1} )^{1 - \alpha } ( {1 - \gamma _{1,d - 1}^ -  } )^\alpha  } ]}}{{1 - \alpha }},
\eeqa
\\
\beqa\label{q13}
R_U ( c ) = \frac{{\log [ {( {\gamma _{1,d - 1}^ +  } )^\alpha   + ( {d - 1} )^{1 - \alpha } ( {1 - \gamma _{1,d - 1}^ +  } )^\alpha  } ]}}{{1 - \alpha }}.
\eeqa

(iii)
When $\alpha = 2$, these lower and upper bounds give the same value.

We use the denotation co($g$) to be the convex hull of the function $g$, which is the largest convex function that is bounded above by $g$, and ca($g$) to be the smallest concave function that is bounded below by $g$. Using the results presented in Methods, we can prove the main result of this paper.

\emph{Theorem.} For any $m \otimes n(m \le n)$ mixed quantum state $\rho$, its ER$\alpha$E satisfies

\beqa\label{q9}
co\left[ {R_L \left( {  \underline C} \right)} \right] \le E_\alpha  \left( \rho  \right) \le ca\left[ {R_U \left( {\overline C} \right)} \right],
\eeqa
where
\beq
\overline C = \min
\bigg\{
{\sqrt {2( {1 - \tr\rho _A^2 } )} ,\sqrt {2({1  - \tr\rho _B^2 } )} }
\bigg\},
\eeq
and
\bea
\underline
C^2 = \max \bigg\{
0,~~2/m(m - 1)( {\left\| {\rho ^{T_A } } \right\| - 1} )^2,
2/m(m - 1) ( {\left\| {\mathcal{R}\left( \rho  \right)} \right\| - 1})^2,
{2\left( {\tr\rho ^2  - \tr\rho _A^2 } \right)},
~~{2\left( {\tr\rho ^2  - \tr\rho _B^2 } \right)}
\bigg\}.
\eea

\begin{figure}[htb]
\includegraphics[scale=0.65,angle=0]{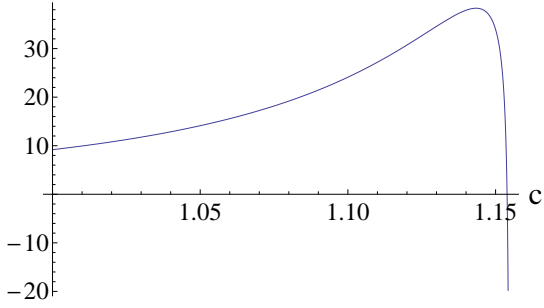}
\caption{(color online). The plot of the second derivative of $R_{12}$ for $1 < c \le 2/{\sqrt 3}$.}\label{fig2}
\end{figure}

Next we consider how to calculate the expressions of $co\left( {R_L \left( {c } \right)} \right)$ and $ca\left( {R_U \left( {c } \right)} \right)$. As an example, we only consider the case $m =3$. In order to obtain $co\left( {R_L \left( {c } \right)} \right)$, we need to find the largest convex function which bounded above by $R_L \left( {c } \right)$. We first set the parameter $\alpha  =3$, then we can derive

\beqa\label{q21}
R_L \left( c \right) = \left\{ {\begin{array}{*{20}l}
   {R_{11} ,0 < c \le 1}  \\
   {R_{12} ,1 < c \le 2/{\sqrt 3 }} , \\
\end{array}} \right.\nonumber\\
R_U \left( c \right) = \left\{ {\begin{array}{*{20}l}
   {R_{11} ,0 < c \le 1}  \\
   {R_{21} ,1 < c \le 2/{\sqrt 3 }}.
\end{array}} \right.
\eeqa

\begin{figure}[htb]
\includegraphics[scale=0.7,angle=0]{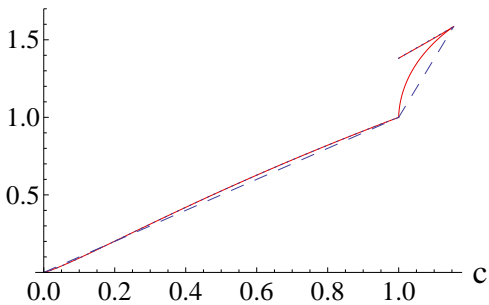}
\caption{(color online). The plot of lower bound (dashed line) and upper bound (dotted line) for $\alpha=0.6, m=3$. The upper bound consists of two segments and the lower bound also consists of two segments. The solid line corresponds to $R_{11},R_{12}$ and $R_{21}$.}\label{fig3}
\end{figure}

We plot the function $R_{11},R_{12}$ and $R_{21}$ in Fig.\ref{fig1} which illustrates our result. It is direct to check that $R''_{11}  \ge 0$, therefore $co\left( {R_{11} } \right) = R_{11}$ for $0 < c \le 1$. The second derivative of $R_{12}$ is not convex near ${c = 2/\sqrt 3 }$ as shown in Fig.\ref{fig2}. In order to calculate $co( {R_{12} } )$, we suppose $l_1\left( c \right) = k_1\left( {c - 2/\sqrt 3 } \right) + \log 3$ to be the line crossing through the point $[2/{\sqrt 3},R_{12}(2/{\sqrt 3})]$. Then we solve the equations $l_1\left( c \right) = R_{12}(c)$ and $dl_1\left( c \right)/dc = dR_{12}(c) /dc = k_1$ and the solution is $k_1=5.2401, c=1.1533$. Combining the above results, we get

\beqa\label{q22}
co\left( {R_L \left( c \right)} \right) = \left\{ {\begin{array}{*{20}l}
   {R_{11} (0 < c \le 1)}  \\
 R_{12} (1 < c \le 1.1533) \\
 5.2401\left( {c - 2/\sqrt 3 } \right) + \log 3 \\
 (1.1533 < c \le 2/\sqrt 3 ). \\
\end{array}} \right.
\eeqa
Similarly, we can calculate that $R''_{11}  \ge 0$ and $R''_{21}  \ge 0$, thus $ca\left( {R_U \left( {c } \right)} \right)$ is the broken line connecting the following points: $[0,0],[1,\log 2],[2/{\sqrt 3},\log 3]$. In Fig.\ref{fig3} we have plotted the lower and upper bounds with dashed and dotted line respectively.

Then we choose the parameter $\alpha  =0.6$, and we get

\beqa\label{q23}
R_L \left( c \right) = \left\{ {\begin{array}{*{20}l}
   {R_{11} (0 < c \le 1)}  \\
   {R_{21} (1 < c \le 2/\sqrt 3 )}, \\
\end{array}} \right.
\eeqa
\beqa\label{q24}
R_U \left( c \right) = \left\{ {\begin{array}{*{20}l}
   {R_{11} (0 < c \le 1)}  \\
   {R_{12} (1 < c \le 2/\sqrt 3 )}.\\
\end{array}} \right.
\eeqa

Since $R''_{11}  \leq 0$, $R''_{21}  \leq 0$, we have that $co\left( {R_L \left( {c } \right)} \right)$ is the broken line connecting the points: $[0,0],[1,\log 2],[2/{\sqrt 3},\log 3]$. In order to obtain $ca\left( {R_U \left( {c } \right)} \right)$, we need to find the smallest concave function which bounded below by $R_U \left( {c } \right)$. We find $R''_{11}  \leq 0$, $R''_{12}  \geq 0$, therefore $ca\left( {R_U \left( {c } \right)} \right)$ is the curve consisting $R_{11}$ for $0 < c \le 1$ and the line connecting points $[1,R _{12} (1)]$ and $[2/\sqrt 3,R _{12} (2/\sqrt 3)]$ for $1 < c \le 2/\sqrt 3$.
As shown in Fig.\ref{fig3}, the lower and upper bound both consists of two segments in this case.

Generally, we can get the expression of $co\left( {R_L \left( {c } \right)} \right)$ and $ca\left( {R_U \left( {c } \right)} \right)$ for other parameters $\alpha$ and $m$ using similar method.

\section*{examples}
\label{sec:ex}

In the following, we give two examples as applications of the above results.

\emph{Example 1}. We consider the $d\otimes d$ Werner states
\beqa\label{q25}
\rho _f  = \frac{1}{{d^3  - d}}\left[ {\left( {d - f} \right)I + \left( {df - 1} \right)\mathcal{F}} \right],
\eeqa
where $- 1 \le f \le 1$ and $\mathcal{F}$ is the flip operator defined by ${\mathcal{F}\left( {\phi  \otimes \psi } \right) = \psi  \otimes \phi }$. It is shown in Ref. \cite{Chen06rmp} that the concurrence $C\left( {\rho _f } \right) =  - f$ for $f<0$ and $C\left( {\rho _f } \right) =  0$ for $f\geq 0$. According to the theorem we obtain that
\\$1/(1 - \alpha )\log \left[ {\left( {(1 + \sqrt {1 - f^2 } )/2} \right)^\alpha  + \left( {(1 - \sqrt {1 - f^2 } )/2} \right)^\alpha  } \right] \le $ $
E_\alpha  \left( {\rho _f } \right) \le  - f$ for $- 1 \le f \le 0$ when $m=3$.

\emph{Example 2}. The second example is the ${3 \otimes 3}$ isotropic state $\rho  = (x/9) I + \left( {1 - x} \right)\left| \psi  \right\rangle \left\langle \psi  \right|$, where \\$\left| \psi  \right\rangle  = \left( {a,0,0,0,1/\sqrt 3 ,0,0,0,1/\sqrt 3 } \right)^t /\sqrt {a^2  + 2/3}$ with $0 \le a \le 1$. We choose $x=0.1$, it is direct to calculate that

\beqa\label{q26}
  C_1 = \sqrt {2\left( {Tr\rho ^2  - Tr\rho _A^2 } \right)}= \sqrt {2\left( {Tr\rho ^2  - Tr\rho _B^2 } \right)}
  = \frac{{2\sqrt {6.53 + 41.46a^2  - 1.71a^4 } }}{{3\left( {2 + 3a^2 } \right)}},
\eeqa

\beqa\label{q27}
 C_2  = \frac{1}{{\sqrt 3 }}\left( {\left\| {\rho ^{T_A } } \right\| - 1} \right)
  = \frac{{2\left( {5 + 6.9a^2  - 0.9a^4  + 9.353a(2 + 3a^2 )} \right)}}{{3(2 + 3a^2 )^2 }},
\eeqa

\beqa\label{q28}
 C_3  = \frac{1}{{\sqrt 3 }}\left( {\left\| {R\left( \rho  \right)} \right\| - 1} \right) = \frac{{{\rm{0}}{\rm{.346  + 1}}{\rm{.2a}}}}{{{\rm{0}}{\rm{.667 + }}a^2 }},
\eeqa

\beqa\label{q29}
 \overline C = \sqrt {2\left( {1 - Tr\rho _A^2 } \right)}= \sqrt {2\left( {1 - Tr\rho _B^2 } \right)}
  = \frac{{\sqrt {6(6.38 + 33.72a^2  + 3.42a^4 )} }}{{3 \left( {2 + 3a^2 } \right)}}.
\eeqa

\begin{figure}[htb]
\includegraphics[scale=0.7,angle=0]{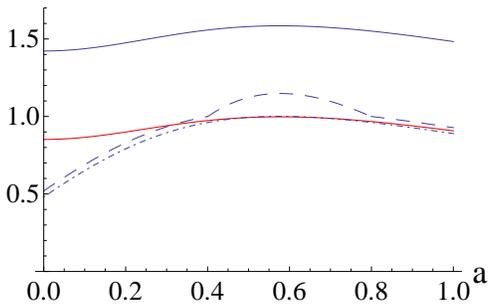}
\caption{(color online). Lower and upper bounds of $E_\alpha  \left( \rho  \right)$ for $\alpha=0.6$ where we have set $x=0.1$. Red solid line is obtained by $C_1$, the dash-dotted and dashed line is obtained by $C_2$ and $C_3$, respectively. The blue solid line is the upper bound of $E_\alpha  \left( \rho  \right)$.}\label{fig4}
\end{figure}

When $\alpha=0.6$, we can calculate the lower and upper bounds and the results is shown in Fig.\ref{fig4}. The solid red line corresponds to the lower bound of $E_\alpha$ by choosing the lower bound of concurrence is $C_1$, and the dash-dotted and dashed line correspond to the cases when we choose the lower bound of concurrence is $C_2$ and $C_3$, respectively. We can choose the maximum value of the three curves as the lower bound of $E_\alpha$. The blue solid line is the upper bound of $E_\alpha$.

\section*{relation with other entanglement measures}
\label{sec:co}

In this section we establish the relation between ER$\a$E and other well-known entanglement measures, such as the entanglement of formation, the geometric measure of entanglement \cite{wg03}, the logarithmic negativity and the G-concurrence.

\subsection*{entanglement of formation}
\label{sec:eof}

Let $\r$ be a bipartite pure state with Schmidt coefficients $(\m_1,\m_2,\cdots)$. We investigate the derivative of ER$\a$E w.r.t. $\a$ as follows.
\bea\label{q30}
{d E_\alpha (\r) \over d \a}
&=&
{1\over(1-\a)^2}
\bigg(
\sum_j {\m_j^\a \over \sum_k \m_k^\a} \log \m_j^{1-\a}
+
\log  \sum_k \m_k^\a
\bigg)
\notag\\
&\le&
{1\over(1-\a)^2}
\bigg(
 \log{ \sum_j \m_j \over \sum_k \m_k^\a}
+
\log  \sum_k \m_k^\a
\bigg)
\notag\\
&=&0.
\eea
The inequality follows from the concavity of logarithm function. The last equality follows from the fact $ \sum_j \m_j =1$. Hence the ER$\a$E is monotonically non-increasing with $\a\ge0$. Since it becomes the von Neumann entropy when $\a$ tends to one, we have
\bea\label{q31}
E_\a(\r)\ge E_F(\r) \ge E_{\b}(\r)
\eea
where $0\le\a\le1$ and $\b\ge1$. Using the convex roof, one can show that \eqref{q31} also holds for mixed bipartite states $\r$.

\subsection*{geometric measure of entanglement}
\label{sec:gm}

The geometric measure (GM) of entanglement measures the closest distance between a quantum state and the set of separable states \cite{wg03}. The GM has many operational interpretations, such as the usability of initial states for Grovers algorithm, the discrimination of quantum states under LOCC and the additivity and output purity of quantum channels, see the introduction of \cite{Chen14pra} for a recent review on GM. For pure state $\ket{\ps}$ we define $\Gl(\ps)=-\log \max \abs{\braket{\ph}{\ps}}^2$, where the maximum runs over all product states $\ket{\ph}$. it is easy to see that $\max \abs{\braket{\ph}{\ps}}^2$ is equal to the square of the maximum of Schmidt coefficients of $\ket{\ps}$. For mixed states $\r$ we define

\begin{eqnarray}\label{q32}
    \Gcl ( \rho ) := \min  \sum_{i} p_{i} \Gl \left( \ket{\psi_{i}} \right) \, ,
  \end{eqnarray}
where the minimum runs over all decompositions of $\rho = \sum_{i}
p_{i} \proj{\psi_{i}}$ \cite{Chen14pra}. We construct the linear relation between the GM and ER$\a$E as follows.

\textit{Lemma }.
\label{le:gm}
If $\a>1$ then
\begin{eqnarray}
\label{le:gma>1}
{\a\over2(\a-1)}\Gcl(\r)\ge E_\alpha (\r).
\end{eqnarray}
If $\a=1$ and $\r$ is a pure state then
\begin{eqnarray}
\label{le:gma=1}
\Gcl(\r)\le E_\alpha (\r).
\end{eqnarray}
If $\a<1$ then
\begin{eqnarray}
\label{le:gma<1}
E_{\a}(\r)+{\a\over2(1-\a)}\Gcl(\r) \le {1\over1-\a}\log d,
\end{eqnarray}
where $d$ is the minimum dimension of $\cH_A$ and $\cH_B$. The details for proving the lemma can be seen from Methods.

\subsection*{logarithmic negativity}

In this subsection we consider the logarithmic negativity \cite{Audenaert03prl}. It is the lower bound of the PPT entanglement cost \cite{Audenaert03prl}, and an entanglement monotone both under general LOCC and
PPT operations \cite{Plenio05prl}. The logarithmic negativity is defined  as

\beqa\label{q34}
LN(\rho ) = \log  \left\| {\rho ^{T_A } } \right\|.
\eeqa

Suppose $\rho  = \sum\nolimits_i {p_i \left| {\psi _i } \right\rangle \left\langle {\psi _i } \right|}$ is the optimal decomposition of ER$\alpha$E $E_\alpha  \left( \rho  \right)$, and the pure state $\ket{\ps_i}$ has the standard Schmidt form $\ket{\ps_i}=\sum_j \sqrt{\m_{i,j}} \ket{a_{i,j},b_{i,j}}$. For $1/2\leq\alpha\leq (2n-1)/2n$ and $n>1$, we have

\begin{eqnarray}\label{q35}
 n\times LN(\rho ) &=& n\log  \left\| {\rho ^{T_A } } \right\|
  \ge n\log  \sum\limits_i {p_i } \left\| {\left( {\left| {\psi _i } \right\rangle \left\langle {\psi _i } \right|} \right)^{T_A } } \right\|
  \ge n\sum\limits_i {p_i } \log  \left\| {\left( {\left| {\psi _i } \right\rangle \left\langle {\psi _i } \right|} \right)^{T_A } } \right\| \notag\\
  &=& 2n\sum\limits_i {p_i } \log  ( {\sum\limits_j {\sqrt {\mu _{i,j} } } } )
    \ge 2n\sum\limits_i {p_i } \log  \sum\limits_j {\mu _{i,j}^\alpha  }
  \ge \frac{1}{{1-\alpha}}\sum\limits_i {p_i } \log  \sum\limits_j {\mu _{i,j}^\alpha  }  \notag\\
  &=&   \sum\limits_i {p_i } E_\alpha  \left( {\left| {\psi _i } \right\rangle } \right)
  =   E_\alpha  \left( \rho  \right)
\end{eqnarray}
where the first inequality is due to the property proved in \cite{Plenio05prl}, the second inequality is due to the concavity of logarithm function, and in the last inequality we have used the inequality $2n\geq 1/(1-\alpha)$ for $1/2\leq\alpha\leq (2n-1)/2n, n\geq1$.

\subsection*{G-concurrence}

The G-concurrence is one of the generalizations of concurrence to higher dimensional case. It can be interpreted operationally as a kind of entanglement capacity \cite{Gour05pra,Gour05pra2}. It has been shown that the G-concurrence plays a crucial role in calculating the average entanglement of random bipartite pure states \cite{Cappellini06pra} and demonstration of an asymmetry of quantum correlations \cite{Horodecki06qh}. Let $\ket{\ps}$ be a pure bipartite state with the Schmidt decomposition $\ket{\ps}=\sum^d_{i=1}\sqrt{\m_i} \ket{ii}$. The G-concurrence is defined as the geometric mean of the Schmidt coefficients \cite{Gour05pra,Gour05pra2}

\beqa\label{q36}
G\left( {\left| \psi  \right\rangle } \right): = d\left( {\mu _1 \mu _2  \cdots \mu _d } \right)^{1/d}.
\eeqa

For $\alpha > 1$, we have
\begin{eqnarray}\label{q37}
 E_\alpha  \left( {\left| \psi  \right\rangle }  \right) &=& \frac{1}{{1 - \alpha }}\log \sum\limits_i {\mu _i^\alpha  }  \notag\\
  &\le& \frac{1}{{1 - \alpha }}\log \left( {d\left( {\mu _1^\alpha   \cdots \mu _d^\alpha  } \right)^{\frac{1}{d}} } \right) \notag\\
  &=& \frac{1}{{\left( {1 - \alpha } \right)}}[\alpha \log d + \log \left( {\mu _{1}  \cdots \mu _{d} } \right)^{\frac{\alpha }{d}}   \notag\\
  &-& \left( {\alpha  - 1} \right)\log d] \notag\\
  &=& \frac{\alpha }{{1 - \alpha }}\log G\left( {\left| \psi  \right\rangle } \right) + \log d.
\end{eqnarray}

For $0 < \alpha < 1$, we have
\begin{eqnarray}\label{q38}
E_\alpha  \left( {\left| \psi  \right\rangle }  \right) \ge \frac{\alpha }{{1 - \alpha }}\log G\left( {\left| \psi  \right\rangle } \right) + \log d.
\end{eqnarray}

\section*{Discussion and conclusion}
\label{sec:dis}

Entanglement R\'{e}nyi-$\alpha$ entropy is an important generalization of the entanglement of formation, and it reduces to the standard entanglement of formation when $\alpha$ approaches to 1. Recently, it has been proved \cite{Song16pra} that the squared ER$\alpha$E obeys a general monogamy inequality in an arbitrary $N$-qubit mixed state. Correspondingly, we can construct the multipartite entanglement indicators in terms of ER$\alpha$E which still work well even when the indicators based on the concurrence and EOF lose their efficacy. However, the difficulties in minimization procedures restrict the application of ER$\alpha$E. In this work, we present the first lower and upper bounds for the ER$\alpha$E of arbitrary dimensional bipartite quantum systems based on concurrence, and these results might provide an alternative method to investigate the monogamy relation in high-dimensional states. We also demonstrate the application our bound for some examples. Furthermore, we establish the relation between ER$\alpha$E and some other entanglement measures. These lower and upper bounds can be further improved for other known bounds of concurrence \cite{Ma11,Vicente07pra}.

\section*{Methods}
\subsection*{Proof of the theorem.}

Suppose $\rho  = \sum\nolimits_j {p_j \left| {\psi _j } \right\rangle \left\langle {\psi _j } \right|}$ is the optimal decomposition of ER$\alpha$E $E_\alpha  \left( \rho  \right)$, and the concurrence of $\left| {\psi _j } \right\rangle$ is denoted as $c_j$. Thus we have
\beqa\label{q19}
 E_\alpha  ( \rho  ) &=& \sum\nolimits_j {p_j E_\alpha  ( {\left| {\psi _j } \right\rangle } )}  = \sum\nolimits_j {p_j H_\alpha  ( {\vec \mu } )}  \nonumber\\
  &\ge& \sum\nolimits_j {p_j co( {R_L ( {c_j } )} )}  \ge co[ {R_L ( {\sum\nolimits_j {p_j c_j } } )} ] \nonumber\\
  &\ge& co[ {R_L ( \underline C )} ],
\eeqa
where the first inequality is due to the definition of $co(g)$; in the second inequality we
have used the monotonically increasing and convex properties of ${co\left( {R_L \left( {c_j } \right)} \right)}$ as a function of concurrence $c_j$; and in the last inequality we have used the lower bound of concurrence. On the other hand, we have

\beqa\label{q20}
 E_\alpha  ( \rho  ) &=& \sum\nolimits_j {p_j E_\alpha  ( {\left| {\psi _j } \right\rangle } )}  = \sum\nolimits_j {p_j H_\alpha  ( {\vec \mu } )}   \nonumber\\
  &\le& \sum\nolimits_j {p_j ca( {R_U ( {c_j } )} )}  \le ca[ {R_U ( {\sum\nolimits_j {p_j c_j } } )} ]  \nonumber\\
  &\le& ca[ {R_U ( \overline C )} ],
\eeqa
where the first inequality is due to the definition of $ca(g)$; the second inequality is due to the monotonically increasing and concave properties of ${ca( {R_U ( {c_j } )} )}$ as a function of concurrence $c_j$; and in the last inequality we have used the upper bound of concurrence. Thus we have completed the proof of the theorem.
\subsection*{Proof of the lemma.}

Suppose the minimum in \eqref{q32} is reached at $\rho = \sum_{i}
p_{i} \proj{\psi_{i}}$. Let the Schmidt decomposition of $\ket{\ps_i}$ be $\ket{\ps_i}=\sum_j \sqrt{\m_{i,j}} \ket{a_{i,j},b_{i,j}}$ where $\m_{i,1}$ is the maximum Schmidt coefficient. For $\alpha>1$, we have

\begin{eqnarray}\label{q33}
{\a\over2(\a-1)}\Gcl(\r)
&=&
-{\a\over2(\a-1)}\sum_{i} p_{i} \log \m_{i,1}^2
\notag\\
&=&
-{1\over \a-1}\sum_{i} p_{i} \log \m_{i,1}^{\a}
\notag\\
&\geq&
-{1\over \a-1}\sum_{i} p_{i} \log (\sum_j \m_{i,j}^{\a})
\notag\\
&=&
\sum_i p_i E_\alpha(\ket{\psi _i })
\notag\\
&\ge&
E_\alpha (\r).
\end{eqnarray}
We have proved \eqref{le:gma>1}. For $\a=1$, let $\m_i$ be the Schmidt coefficients of $\r$, we have
\begin{eqnarray}\label{q332}
&&E_{\a}(\r)=S(\r)=-\sum_i \m_i \log \m_i
\notag\\
&\ge& -\sum_i \m_i \log \max_j \{\m_j\}
= - \log \max_j \{\m_j\}
\notag\\
&=&\Gcl(\r).
\end{eqnarray}
We have proved \eqref{le:gma=1}.
For $\a<1$, we have
\begin{eqnarray}\label{q333}
&&
E_{\a}(\r)+{\a\over2(1-\a)}\Gcl(\r)
\notag\\
&=&
E_{\a}(\r)-{\a\over2(1-\a)}\sum_{i} p_{i} \log \m_{i,1}^2
\notag\\
&=&
E_{\a}(\r)-{1\over1-\a}\sum_{i} p_{i} \log \m_{i,1}^{\a}
\notag\\
&=&
E_{\a}(\r)-{1\over1-\a}(\sum_{i} p_{i} \log (d\m_{i,1}^{\a}) - \log d)
\notag\\
&\le&
E_{\a}(\r)-{1\over1-\a}\sum_{i} p_{i} \log (\sum_j \m_{i,j}^{\a})
+{1\over1-\a}\log d
\notag\\
&\le&{1\over1-\a}\log d.
\end{eqnarray}
The inequality holds because the pure state $\ket{\psi_{i}}$ is in the $d\times d$ space. So we have proved \eqref{le:gma<1}.

\section*{Acknowledgements}
After completing this manuscript, we became
aware of a recently related paper by Leditzky \emph{et al}\cite{Leditzky11} in which they also obtained another lower bound of ER$\alpha$E in terms of R\'{e}nyi conditional entropy. WS was supported by NSF-China under Grant Nos.11374085, 11274010, the discipline top-notch talents Foundation of Anhui Provincial Universities, the Excellent Young Talents Support Plan of Anhui Provincial Universities, the Anhui Provincial Natural Science Foundation and the 136 Foundation of Hefei Normal University under Grant
No.2014136KJB04. LC was supported by the NSF-China (Grant No. 11501024), and the Fundamental Research Funds for the Central Universities (Grant Nos. 30426401, 30458601 and 29816133).

\section*{Author contributions statement}

 W. Song and L. Chen carried out the calculations. W. Song and L. Chen conceived the idea. All authors contributed to the interpretation of the results and the writing of the manuscript. All authors reviewed the manuscript.

\section*{Additional information}

\textbf{Competing financial interests:} The author declares no competing financial interests.

\end{document}